# An LHC Run Plan –
# The First Inverse Femtobarn

**Dan Green**
**Fermilab**

**January, 2006**

**Introduction:**

The LHC will open up a new regime in high energy physics with a factor seven in C.M. energy increase over the Tevatron and an approximate factor of one hundred in luminosity when the LHC reaches the luminosity design value. Since those increases imply a greatly expanded discovery reach, it is incumbent on the LHC experimenters to be prepared for discoveries, even in the very early days of operation at much lower luminosities.

Manufacturing and survey data will give the tracking detectors of an LHC experiment an initial, though crude, alignment. Cosmic ray muons can be used to pre-align the tracking detectors, both the central tracking and the muon chambers themselves. Shifts in detectors due to magnetic forces arising from powering the analysis magnet(s) must also be tracked and corrected for. Of course, ultimately, tracks themselves will be used to make the final alignments, but a first pass alignment is needed, even to find tracks at all. The muon systems at the LHC can typically not trigger efficiently on high $P_T$ muons without a sufficiently accurate pre-alignment.

Finally the calorimeters need to be pre-calibrated. Using manufacturing data, test beams, radioactive sources, and cosmic ray muons an initial calibration to 5% accuracy seems quite plausible. This initial crude calibration can be refined using several reaction channels. In addition, the overall energy scale can also be set using the fact that top pairs are copiously produced at the LHC. At each decade increase in luminosity, as the LHC starts operation, new channels to use in calibration become useful, and redundant checks can be made.

During the pre-operations phase before colliding beams become available, preparations should also be made. Even before collisions, single beam data and muon halo will be useful in aligning muon and central tracking chambers. The synchronization of all the sub-detectors of a LHC detector with a 40 MHz clock is a formidable task. It can be confronted by using electronic pulse injection and/or laser signal injection. These methods should allow a given sub-detector to be relatively timed to about 1 nsec for all the channels of that subsystem.

**The First 5 Orders of Magnitude:**

When the LHC begins operation, useful data can be taken for more than ten orders of magnitude in luminosity with respect to the design level. That fact argues most strongly that the LHC experimenter be ready, as these many decades of rate will most likely come very quickly in the beginning. This note covers the vast luminosity range in two steps. First there are five orders of magnitude where minimum bias and jet events are initially logged. The second section covers the next six orders of magnitude where processes with single and di-leptons become evident along with single and di-photons.

A summary of the run plan for the first five orders of magnitude of useful LHC luminosity is shown in Table 1. The entries are the instantaneous luminosity for one month, the integrated luminosity, the trigger conditions, the basic reactions of interest, and the comments on the run plan at that stage of luminosity evolution.

**Table 1: Run plan summary for the first five orders of magnitude of useful LHC luminosity**

| L for 1 month run ($10^6$ sec) | Integrated L | Trigger | Process | Comments |
|---|---|---|---|---|
| $10^{23}$ | 100 mb$^{-1}$ | None $\sigma_I \sim$ 50 mb | Inelastic non-diff | Input to tweak Pythia |
| $10^{24}$ | 1 μb$^{-1}$ | Setup Jet | Inelastic non-diff | Calib in azimuth |
| $10^{25}$ | 10 μb$^{-1}$ | Jet $\sigma(gg) \sim$ 90 μb $\sigma(ggg) \sim$ 6 μb | g+g -> g+g g+g -> g+g+g | Establish JJ cross section |
| $10^{26}$ | 100 μb$^{-1}$ | Jet | g+g -> g+g g+g -> g+g+g | Dijet balance for polar angle – Establish MET |
| $10^{27}$ | 1 nb$^{-1}$ | Jet Setup Photon $\sigma(q\gamma) \sim$ 20 nb | g+g -> g+g g+g -> g+g+g q+g -> q+γ | Dijet masses > 2 TeV, start discovery search. J+γ calib |

Note that the total non-diffractive inelastic rate varies from 0.005 Hz to 50 Hz over the range of luminosity shown in Table 1. Therefore, all events can be logged without bias if the LHC experiment has a sufficiently rapid data acquisition system. If this is the case, then the experimenters can study jet algorithms with an unbiased sample, determine the trigger and reconstruction efficiencies for jets, and set up the jet triggers to be used near the end of this first period.

At the end of this period, 1 nb$^{-1}$, there will be 50 million inelastic events logged and 90,000 jets. The jet rate will be about 0.09 Hz. What can be done with these events? First, let us look at the minimum bias inelastic events. There is no fundamental theory available to model these events. Indeed as shown in Fig.1 and Fig.2 there is a substantial extrapolation from the Tevatron to the LHC in regards to the particle density on the rapidity plateau and the mean transverse momentum per particle respectively.

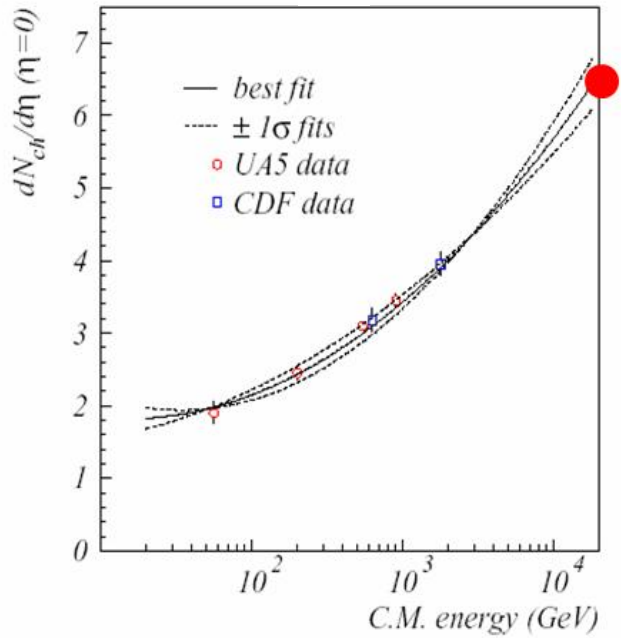

Figure 1: Data from lower energies used to extrapolate the charged particle rapidity density to the LHC. The dot shows a typical resulting estimate.

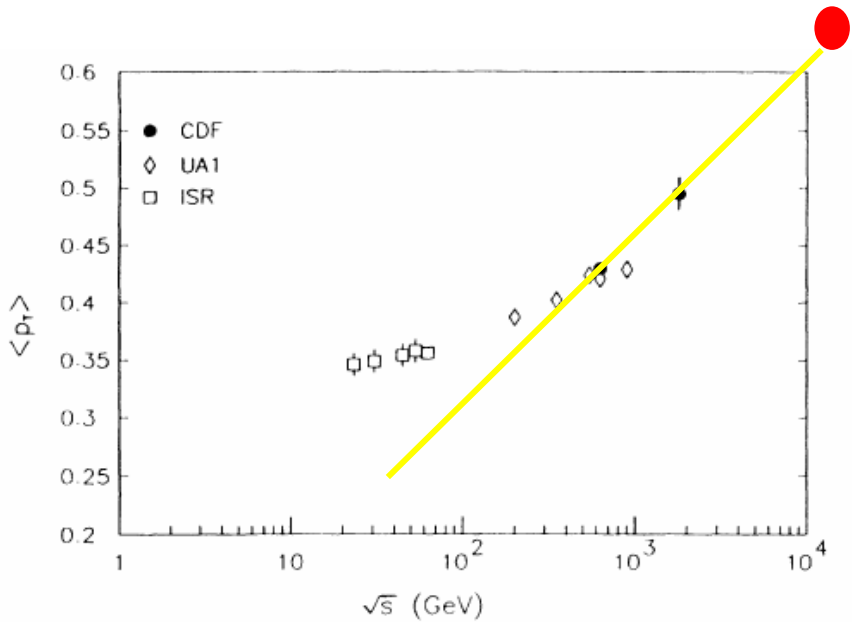

Figure 2: Data from lower energies used to extrapolate the mean transverse momentum of charged particles to the LHC. The dots show a typical resulting estimate.

A plausible value might be nine particles per unit of rapidity with a mean $P_T$ of 0.65 GeV, but these values need to be confirmed. There have been Tevatron studies [1] which also explore the underlying event in di-jet and Drell-Yan events. Input from these processes can be used to refine the Monte Carlo models which model the minimum bias events [2].

Clearly, the Monte Carlo models need to be "tweaked" to better represent the data actually observed. This is particularly so, because at the LHC the details of the inelastic events are crucial to strategies for imposing isolation on leptons and photons, for dealing with pileup, and for trigger algorithms. These strategies must be quickly re-evaluated in the light of the data first taken at the LHC.

Since the p-p system is azimuthally symmetric, that property can be invoked to relatively calibrate all the identical calorimeter towers in a rapidity range. At the end of the 10 $\mu b^{-1}$ exposure, there are 500,000 inelastic events and 30,000 charged particles in a generic calorimeter tower with area $\Delta\eta\Delta\phi/2\pi = (0.1)^2$ which deposit, ignoring magnetic field effects, 19 TeV per tower. For neutrals, there are 15,000 particles in that tower area depositing 9.5 TeV of energy. The rates scale with tower "area" under the assumption that there is a rapidity plateau. These statistics are sufficiently large that one will be systematics limited rather rapidly, for example by azimuthal non-uniformities in the detectors themselves. If the hadronic calorimetry has a fractional energy resolution of $100\%/\sqrt{E_t}$, then the 19 TeV of energy is measured to better than 1%. At the end of the first three orders of magnitude, it can be assumed that all towers at a given rapidity have been equalized in azimuth.

At this point, attention can move to the di-jet events. We start with strong interactions initiated by gluons as these yield the largest cross sections. At 10 $\mu b^{-1}$ there are 900 di-jet events where both jets have transverse momenta above 30 GeV, a typical threshold to find jets efficiently, and are both contained within the acceptance of the experiment, |y|<5. Events were generated with these cuts using the COMPHEP program [3]. As mentioned above, since the experiment can be initially untriggered, the di-jet cross section can be established by using the unbiased events to determine the jet trigger and reconstruction efficiency. The cross section so determined can be compared to Monte Carlo models [2] to check for consistency. The COMPHEP distribution of initial state gluon longitudinal momentum is shown in Fig.3.

Clearly, the kinematic relationship, $x_1 x_2 = M^2/s, <x> \sim M/\sqrt{s}$, is obeyed with the lower limit on the di-gluon final state mass, M, set by the threshold jet cuts of 30 GeV $<P_z> \sim M/2$. The scatter plot of the angles of the outgoing gluons is shown in Fig.4. The di-jet system is seen to be well contained in a typical LHC detector which spans |y| < 5. Note that, for any number of jets in the final state, the final state mass M can be reconstructed and the total final state longitudinal momentum fraction, x, can also be found ( $x_1 - x_2 = x$ ). Given x and M, the initial state momentum fractions, for example see Fig.3, can be solved for and compared to Monte Carlo expectations.

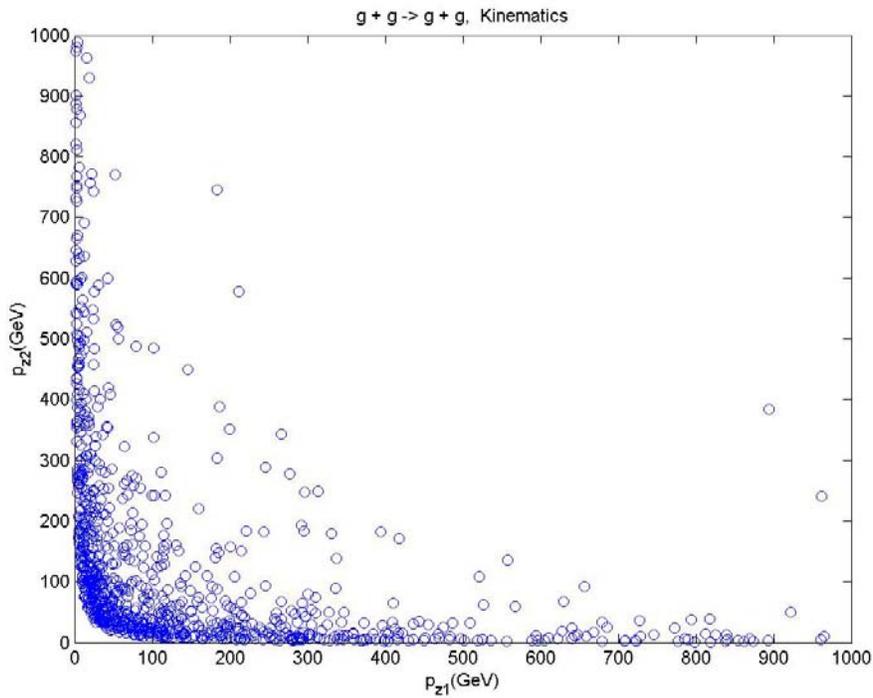

Figure 3: COMPHEP output for the process g + g -> g + g. The scatter plot is the longitudinal momentum of one initial state gluon vs. the other.

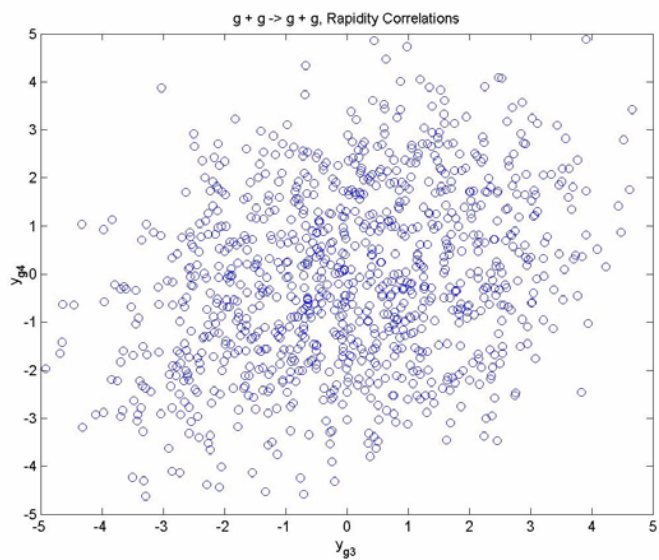

Figure 4: Scatter plot of the rapidity of one outgoing gluon vs. the other the process g + g -> g + g. Note that the plot limits are taken to be the detector acceptance limits.

The transverse momentum distribution of one of the gluons is shown in Fig.5. Note that log-log axes are used. This is done because the basic physics has power law behavior due to the point like nature of the scattering and because the structure functions defining the momentum of the point like constituents in the proton also has a power law behavior. Thus, there is an effective power law behavior, seen in Fig.5, with $d\sigma/dP_T \sim (1/P_T)^5$ which is also shown as a line in the figure. Note that simple point like scattering behavior would yield an inverse third power of $P_T$. The steeper behavior is due to added contributions to the falloff from the structure functions.

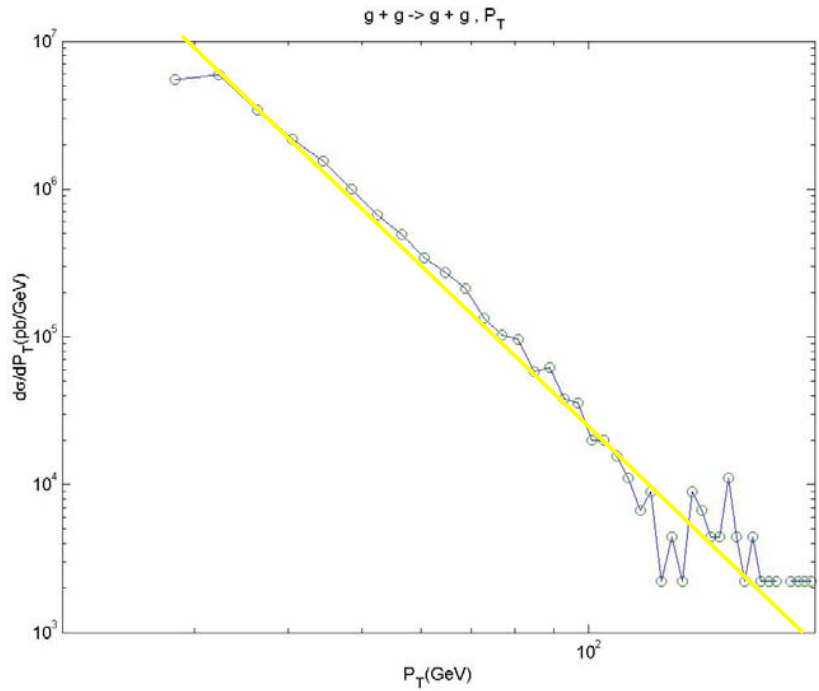

Figure 5: Transverse momentum distribution of one of the gluons in the process g + g -> g + g at the LHC. The line is drawn to indicate the approximate power law behavior.

Another cross check with models is to look at the angular distribution of the scattering as shown in Fig.6. It is expected to mirror the t channel exchange behavior of the fundamental scattering [4]. Testing the experimental result provides a good cross check of the calibration of the calorimetry and the possible systematic errors in the trigger and reconstruction procedures.

The initial state is assumed to have small transverse momentum. Therefore, in the absence of initial state or final state radiation, ISR or FSR, the di-jet final state will have no net transverse momentum. Using this fact, observed at the Tevatron, a calibrated tower can be related to an uncalibrated one by requiring that the jet "balance", i.e. that the quantity $\delta = (P_{T1} - P_{T2})/(P_{T1} + P_{T2})$, be minimized.

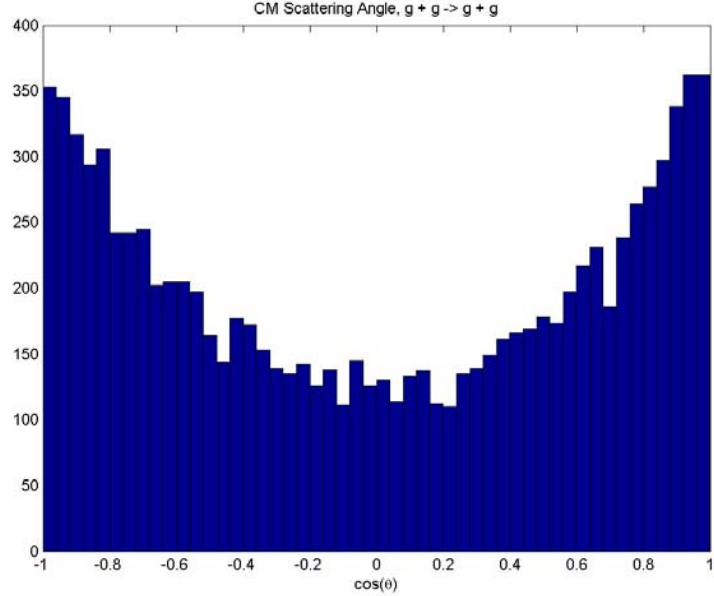

Figure 6: Angular distribution, cosine of the C.M. scattering angle, in the process g + g -> g + g at the LHC.

Finally, the di-jet events can be used to get a first look at the "missing transverse energy" or MET. In the di-jet balance exercise any non-Gaussian "tails" in the balance, or δ, distribution should be tracked down and eliminated. In Fig.7 is shown the distribution in missing transverse energy, MET, for three processes. For the jet processes a Gaussian stochastic smearing of the jet energy with a 100% stochastic coefficient was assumed.

For di-jets the azimuthal angle between the gluon jet axis and the MET will be zero. Therefore, in di-jet events spurious MET due to energy mismeasures can be identified and removed. However, for three or more gluons in the final state this cut is no longer available. The cross section for three explicit hard jets, all with $P_T > 30$ GeV and $|y| < 5$, is only a factor 15 below that of di-jets (see Table 1). Therefore, a MET signal in three jet events with MET not aligned in azimuth to any of the jets will serve to check the known energy resolution of the calibrated calorimetry

For comparison, the cross section for b quark pair production (anti-particles are shown as upper case symbols here) with subsequent semileptonic decay is shown. The cross section is less than that for di-jet production, but if the jet energy errors are Gaussian, the cross over is at about 100 GeV for MET due to real neutrinos in the final state. Clearly, the existence and MET value of this cross over point depends on having only Gaussian errors and having control of the jet energy resolution. In these ways, the LHC experimenters can begin to understand the MET response of their detectors in a rather early data taking period. Note that complications and additional contributions to MET coming from "pileup" of overlapping inelastic events are not considered here.

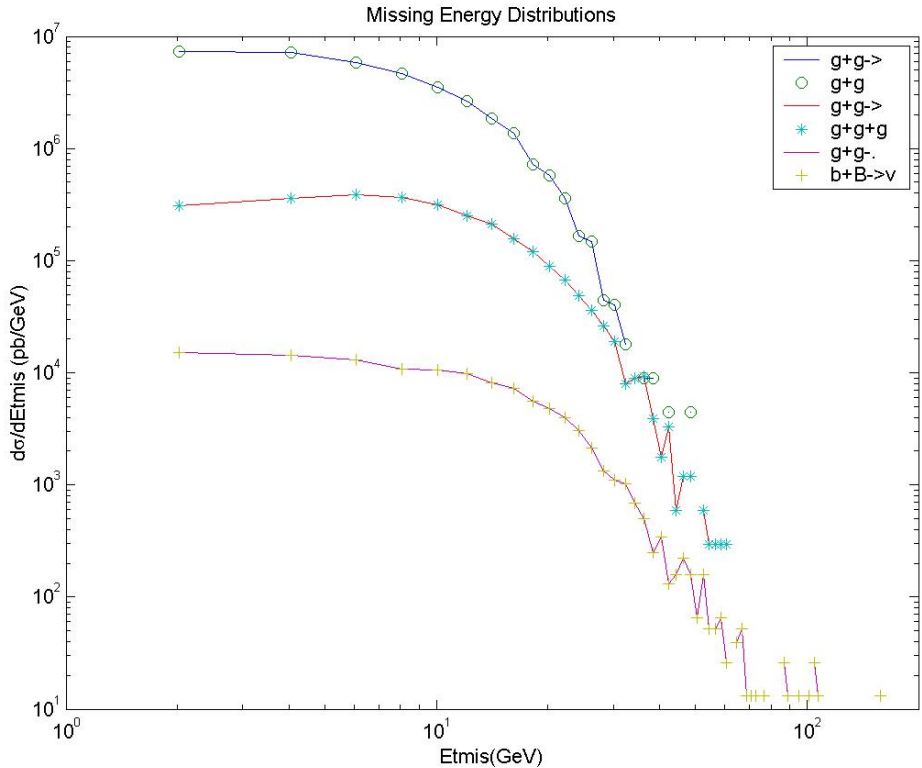

Figure 7: Distribution of MET, or Etmis, for di-jet and tri-jet events and for events due to the process g + g -> b + B with a subsequent decay b -> c + μ+ v. No cut on the azimuthal angle of MET with respect to any jet axis is imposed.

The di-jet events, after they are utilized to check the jet energy resolutions, establish calibrations and understand MET are the first reactions that will take the LHC experimenter into uncharted territory. The two curves seen in Fig.8 are the di-jet mass spectrum for p-p production at 2 and 14 TeV. Clearly, the structure functions at the LHC, being probed at lower x values, allow the experimenters to reach high masses and do so with very low luminosity.

The final state mass distributions for g + g production of g + g, g + g + g, and b + B final states are shown in Fig.9. Note that the log – log plots yield straight line effective power law behavior for all three processes. When the low $P_T$ jets are fully understood exploration of the high mass di-jets can commence. Note that after an exposure of 1 nb$^{-1}$ there will be 90,000 di-jets. For comparison, the number of COMPHEP events in this and most other plots is 10,000.

Clearly, at this point the exploration of the 2 TeV mass scale and beyond can begin. Note that the luminosity is still seven orders of magnitude below the design value, so that this point, especially for a p-p machine, may come very soon in the first running period. As the luminosity increases the gain in mass reach is, very approximately, 1 TeV for each decade in luminosity [5]. Uncharted territory is now accessible.

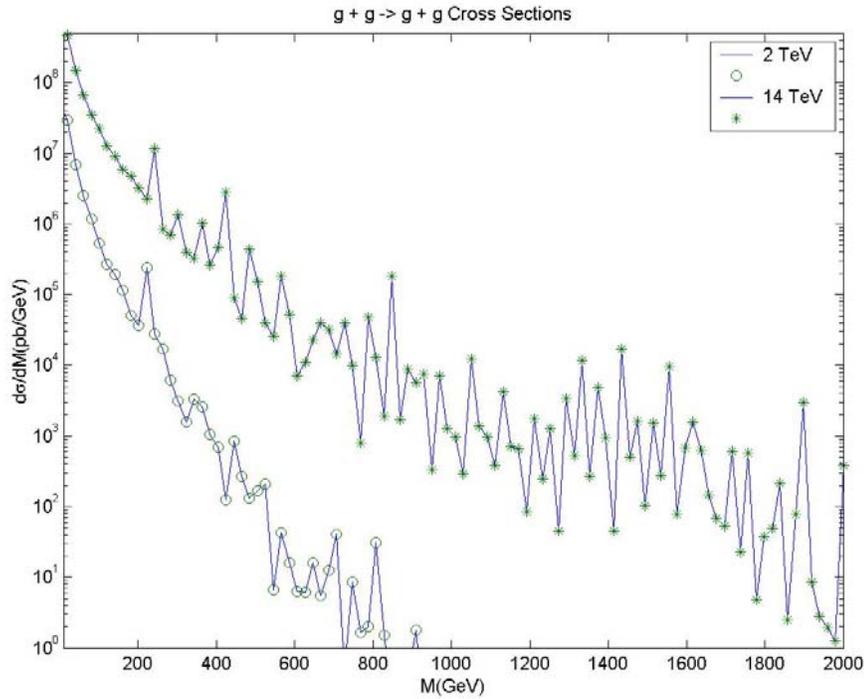

Figure 8: Mass distribution of di-gluons in g + g -> g + g at 2 and 14 TeV C.M. energy. Note the enormous increase in cross section at the LHC with respect to the Tevatron as the mass approaches the Tevatron kinematic limit of 2 TeV.

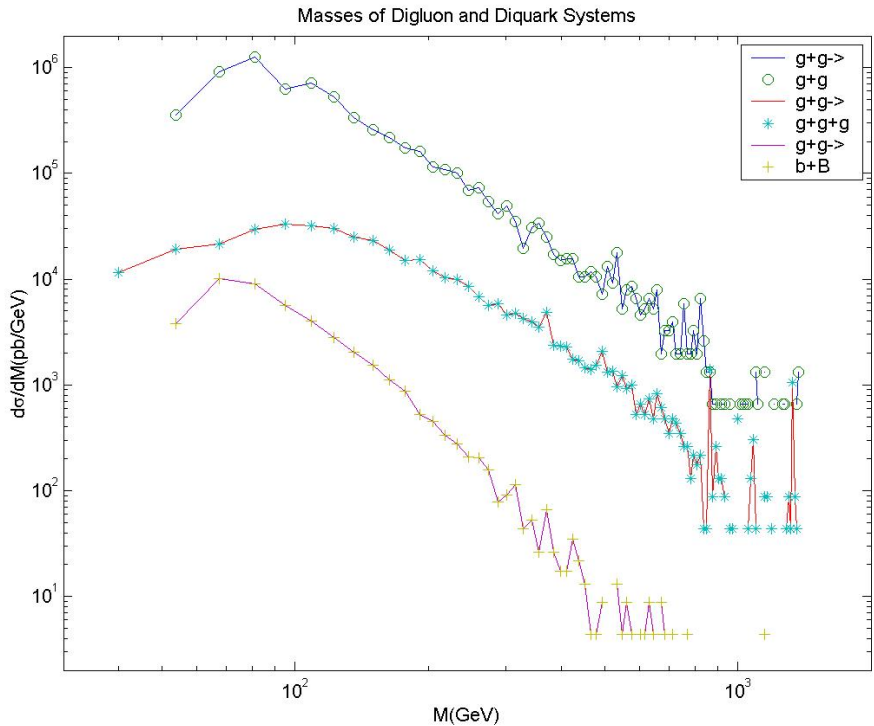

Figure 9: Mass distribution for di-gluon and tri-gluon final states and b quark pair production. There are 10,000 generated events from COMPHEP for each process.

Previously we discussed the use of the di-jets themselves as a calibration tool using di-jet balance. There are other, lower cross section, processes which have different systematic errors or more precise energy measurements than that for di-jets. The cross section for photons or Zs balancing against jets is shown in Fig.10. Once a sufficient number of these events is available a redundant and more accurate energy measurement will become available. However, it must be noted that the number of events using these rarer processes will always be inferior to the number of di-jets, thus making them useful redundant checks but not the main calibration process at the highest mass scales.

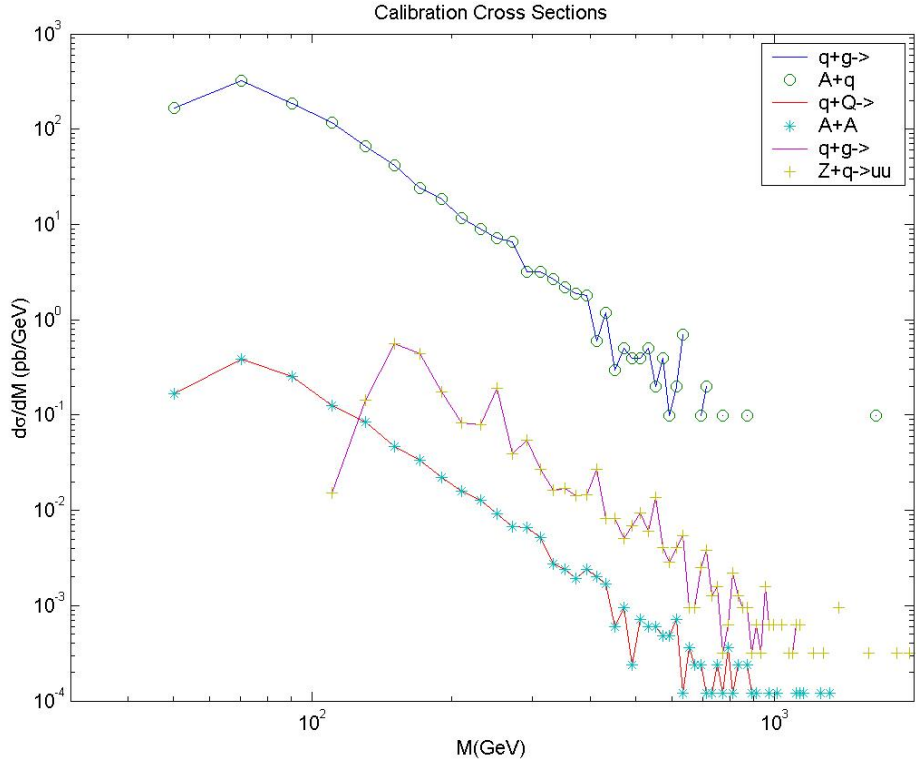

Figure 10: Redundant calibration processes at the LHC. The mass of the 2 body final state for q + photon, q + Z, and di-photon is shown.

The cross section for g + q -> q + $\gamma$ with $P_T$ of the q and photon > 30 GeV and with the photon in the region $|y| < 2.5$ is 20 nb. Therefore, at the end of the first five decades, at 1 nb$^{-1}$, there will be a few events, while at 1 pb$^{-1}$ there will be about 20000 events where an isolated, well measured photon balances a jet in $P_T$. At higher luminosities the g + q -> q + Z process, where the Z is identified in the dimuon final state, will become available as yet another 2 body jet balance process. Note that the cross sections for photon plus jet and Z plus jet are comparable, see Fig.10, at high mass after the di-muon Z decay branching fraction is taken into account. Finally, at still lower cross section the di-photon final state becomes accessible. This process will allow a purely photonic final state which probes only the electromagnetic resolution of the calorimetry. Note that fluctuations of jets to almost all neutrals are also acceptable for this calibration procedure, so that photon cleanliness is not an issue for calibration.

At a final state mass of 100 GeV the di-gluons have a cross section of 600000 pb/GeV (see Fig.9), the b quark pairs 7000 pb/GeV, the photon + jet 200 pb/GeV, and Z + jet ~ 60 pb/GeV. Note that the di-jet to photon+jet ratio is about 3000, so good rejection of spurious photon signals is needed to extract the single photon signal. Establishing a clean isolated single photon signal is a necessary precursor to initiating a di-photon mass search. The ratio of di-gluons to di-photons at 100 GeV mass is about 3 million, so again good jet rejection will be needed to establish the true di-photon signal. The ratio of b quark pairs to di-gluons is (see Fig.9) about one hundred to one, so good b quark tagging, either through secondary vertex separation or soft lepton tags will be needed to isolate b pair production.

For these first five decades of luminosity a detector at the LHC can have the calorimetry calibrated with inelastic events and jets. The inelastic events can be used to model the underlying event. The jet trigger can be set up, the efficiency can be measured, and the triggering can be established. The MET signal due to mismeasures can be checked against the resolution for jets. Finally, the first LHC search can be launched – di-jet masses above 2 TeV. Recall that we are still seven decades away from LHC design luminosity.

**The Next 6 Orders of Magnitude:**

The initial data taking described above largely concerns itself with the modeling of soft interactions, the calibration of the calorimetry (largely with jets), and the study of low $P_T$ jets and missing transverse energy (MET). As the luminosity increases processes with leptons in the final state become available in statistically significant quantities. The next progression by decade of luminosity is given in Table 2.

The first process to be seen will be the production of heavy quark pairs. With very minimal cuts the b pair cross section is about 300 μb which is an enormous cross section, leading to a rate of 3 Hz at a luminosity of only $10^{28}$/cm$^2$*sec.. After requiring that the quark jets be triggerable, $|y| < 5$ and $P_T > 30$ GeV, the cross section for b quark pairs is still 600 nb. That is a factor about 150 less than the cross section for di-gluons. That ratio is understandable as the ratio of the fundamental two body cross sections for g+g-> g+g and g+g->b+B at a scattering angle of 90 degrees [4], which is a factor about two hundred.

The heavy flavors provide an exercise in b tagging using the vertex detectors of the LHC experiments. With an exposure of 10 nb$^{-1}$ there will be about 1200 muonic decays of the b jets and about sixty events with two muonic decays. Note that, Table 1, there will be 900,000 di-jets produced at this point, so that rather good gluon jet rejection will be required to extract a clean b quark pair signal.

**Table 2: Run plan summary for the second six orders of magnitude of useful LHC luminosity**

| L for 1 month run | Integrated L | Trigger | Process | Comments |
|---|---|---|---|---|
| $10^{28}$ | 10 nb$^{-1}$ | $\sigma_{bB}$ ~ 600 nb. Setup – run single electron, muon, photon | g+g -> b+B | 900,000 JJ, 6000 bB, 1200 1μ, 60 2μ Establish μ jet tag |
| $10^{29}$ | 100 nb$^{-1}$ | Setup dimuon, dielectron $\sigma_{\mu\nu}$ ~ 10 nb | q+Q->W->μ+ν (D-Y) | 1000 μ from W->μ + ν Lumi – standard candle (look at high Mt tail) |
| $10^{30}$ | 1 pb$^{-1}$ | Run dilepton trigger $\sigma_{\mu\mu}$~ 1.5 nb $\sigma_{tT}$ ~ 630 pb | q+Q->Z->μ+μ (D-Y) g+g->t+T | 1500 dimuons from Z-mass scale, resolution Lumi- standard candle, high M 600 t + T produced |
| $10^{31}$ | 10 pb$^{-1}$ End of '07 Pilot Run | Setup, J*MET $\sigma_{q\mu\mu}$ ~ 40 pb $\sigma_{\gamma\gamma}$ ~ 24 pb | g+q->Z+q->μ+μ+q q+Q->γ+-γ (tree) | 400 Z + J events with Z->dimuons – Z+J balance, calib Estimate J + MET ( q + ν ) 240 diphoton events with M > 60 GeV |
| $10^{32}$ | 100 pb$^{-1}$ | $\sigma_{qQZ}$ ~ 170 pb $\sigma_{qgZg}$ ~ 32 pb $\sigma_{tT}$ ~ 630 pb | g+g->q+Q+Z g+q->q+g+Z+g | 3000 J+J+Z->νν events, Pt>30 500 J+J+Z->μ+μ events, Pt>30 600 J+J+J+Z->νν events 12000 J+J+J+μ+ν events |
| $10^{33}$ | 1 fb$^{-1}$ (1% of design L for 1 yr) End of '08 Physics Run | | | M of dijet in 120000 top events, W-> μ +ν – set Jet energy scale with W mass. Dimuon mass > 1 TeV, start discovery search, dimuons, diphoton search, SUSY search |

These muons can also be used to begin to establish the muon trigger and to check the alignment of the muon chambers in situ. There will also be roughly two hundred photon + jet events, see Table 1, which can be used for jet calibration and for establishing an isolated photon trigger.

A clean b pair sample would allow for a check on trigger and reconstruction efficiency for leptons in jets, assuming that a double b tag requirement is clean enough to isolate a sample of b quark pairs. In addition, there is, in these events, real MET, not due simply to jet energy mismeasures. Therefore, correlating the tag of the jet by vertex tracking and/or soft leptons with a non-Gaussian tail in the jet balance will be another tool to use in understanding the MET determination.

For the next decade increase in luminosity, the leptons can be used to establish the cross section and other properties of single leptons from W gauge bosons. The leptonic decays can serve as a "standard candle" since the cross section can be accurately extrapolated from the current Tevatron results. The cross section for both W and Z production is expected to rise by a factor about five from 2 TeV C.M. energy to 14 TeV. The cross section at the LHC for |y| < 2.5 and $P_T$ > 15 GeV, typical limits for tracking and triggering on leptons, is about 10 nb. It is expected [6] that the W cross section can be extrapolated with 5% accuracy, thus allowing these events to accurately track the LHC luminosity.

The differential cross section for the transverse momentum of single muons from heavy quarks and from W leptonic decays is shown in Fig.11. The Jacobean peak for two body decays of W in Drell-Yan production is very evident, with a peak cross section comparable to that arising from leptonic decays of b quark pairs. With the requirement of isolation of the lepton and also MET in the event, the W signal should be easily extracted.

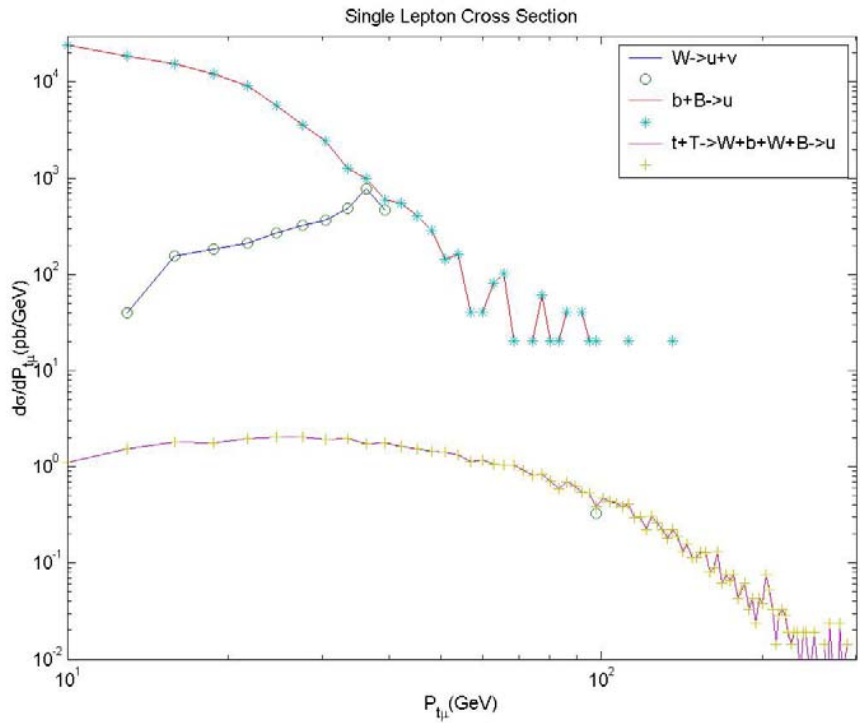

Figure 11: Cross section for single leptons as a function of lepton transverse momentum for W Drell-Yan production and g+g production of heavy b and t quark pairs.

For b quarks the decays are b -> c + μ + v. For the top quarks the decay chain is t -> W + b. The leptons are assumed to come solely from the W -> μ+ v decays. The production cross section at high mass for b and t pairs is the same. However, the intermediate decay, t -> b + W, softens the b spectrum so that the lepton spectrum at high $P_T$ for top has a smaller cross section than that for bottom.

Once the W cross section is established the high tail in the transverse mass distribution can be searched for new phenomena. Again, as with the di-jets, new territory, with transverse masses of leptons plus MET above about 400 GeV, is opened up quite rapidly.

At 1 pb$^{-1}$ of integrated luminosity the di-lepton samples begin to be usable. The Drell-Yan production of Z bosons followed by a di-muon decay has about a 1.5 nb cross section once cuts on both leptons that P$_T$ is > 15 GeV and |y| < 2.5 are made. This process can be used for an in situ calibration where one establishes the mass scale and the mass resolution for di-leptons. Lighter systems such as ψ or Ψ are copiously produced but have final state leptons somewhat below the trigger and reconstruction thresholds. Extrapolation from the Tevatron by a factor about five in cross section also gives the LHC experimenter another "standard candle" with which to normalize the cross sections. The accuracy is also thought to be about 5% for the cross section.

The dimuon mass spectrum is shown in Fig.12. The Z peak is quite prominent even without invoking isolation. The cross sections for heavy quark pair production are related to the single lepton rates of Fig.11 save that now both quarks in the b are required to decay semileptonically or both W from the top decays into W + b are required to decay to μ+ ν.

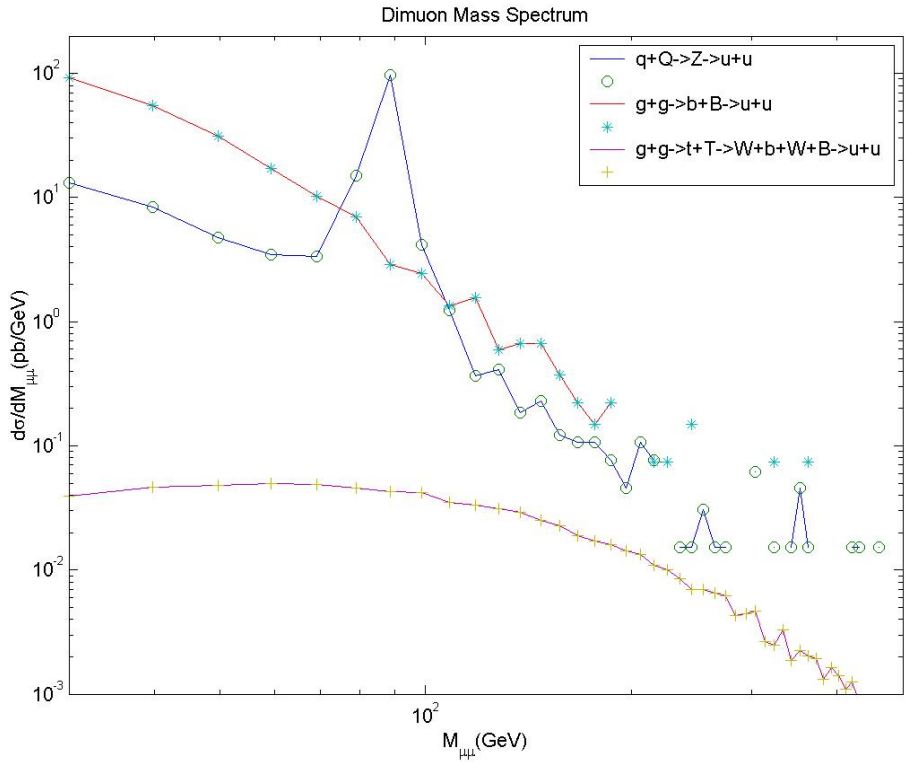

Figure 12: Dimuon mass spectrum for Drell-Yan production of Z followed by dimuon decays and for the production of b and t quark pairs. The b and B both decay semileptonically, b -> c + μ + ν, while the top decays are assumed to be, t -> W + b, W -> μ + ν.

The top cross section, requiring both top quarks to have $P_T > 30$ GeV and $|y| < 5$, is 630 pb. As with single leptons, the cross section for top and b pairs is the same at high mass, so that the rates are comparable, with top being lower because of the spectra softening effect of the intermediate two body decay, while b decays directly to muons. The top pair cross section is about 100 times larger at the LHC than the Tevatron. This fact makes top a background in many searches, e.g. W + W scattering, so that top production will need to be thoroughly understood before searches can be undertaken with backgrounds that are under control.

Once the absolute cross section, the mass scale, and the mass resolution are confirmed for the Z signal the high mass di-lepton spectrum can be explored. Note that there is a substantial background due to real di-muons from heavy flavor decays. Note also that the di-gluon cross section is about one million times larger, Fig.9, than the di-lepton cross section at 500 GeV di-lepton mass. Therefore jet rejections of > 1000 per jet are needed in the di-electron sample, for example. When the backgrounds are under control, the high mass di-lepton region may be explored. At an integrated luminosity of 10 pb$^{-1}$, the end of the 2007 "Pilot Run" of the LHC there should be 15000 di-muon Z decays with a few events above 400 GeV di-muon mass. At the end of the 2008 "First Physics Run" there should be a handful of candidates out in new territory, with a di-lepton mass 1 TeV and above.

Another tool available with di-lepton decays is the forward/backward asymmetry in the decays. This aspect of the decay angular distribution, due to interference of photon and Z amplitudes, can first be studied at low mass, confirming the known Tevatron results. Extrapolating then to high mass, a constant F/B asymmetry is expected in the absence of new physics.

After a cumulative exposure of 10 pb$^{-1}$ the 2007 "Pilot Run" of the LHC is supposed to be completed. At this point there will be, see Table 2, about four hundred events due to the "Compton scattering" process, q + g -> Z + q (see Fig.10) where the Z decays into two muons. There are also a comparable number of events from Drell-Yan production with ISR of a gluon from the q or Q, q + Q -> Z + g ( 1.7 nb), compared to q + g -> Z + q (1.2 nb). These events can begin to be used in yet a third jet balance calibration technique, with the leptons, by assumption, being so well measured as to not contribute to the energy resolution limitation of the technique. These events can also be used to estimate the Z contribution to the jet plus MET cross section where the Z decays into neutrino pairs instead of two leptons.

Finally, the search for new phenomena can be extended to di-photons. The mass distribution is shown in Fig.13. The photons are required to both have $|y| < 2.5$ and $P_T > 30$ GeV. The cross section is about 24 pb. As mentioned before this cross section is about 3 million times smaller than that for di-gluons, so that good jet rejection is mandatory. The falloff with mass goes approximately as the mass to the power 3.3. At the Tevatron the search in the di-photon spectrum at a total exposure of 0.3 fb$^{-1}$ has a few events at 300 GeV mass. Once a clean di-photon signal is established, the search at high di-photon mass can commence. For the LHC the tree level (COMPHEP does only tree diagrams) cross section above 300 GeV is roughly 1 pb, so that there will be about ten events above 300 GeV at the end of the "Pilot Run",  and about 1000 events ( roughly sixty events above 1 TeV) at the end of the "First Physics Run".

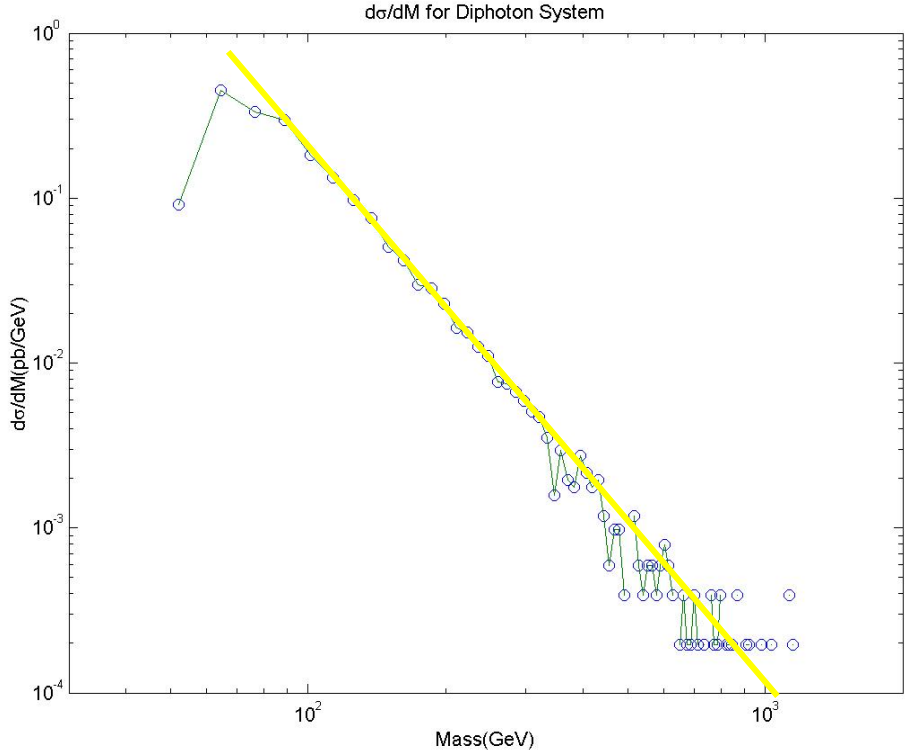

Figure 13: Mass spectrum for di-photons produced in the process q + Q -> γ + γ at the LHC (tree diagrams only).

At an integrated luminosity of 100 pb$^{-1}$ a variety of di-jet , tri-jet, and four jet plus missing energy processes become observable,  The Compton scattering process, g + q  -> q + Z, has a MET cross section for the MET plus one jet topology of around 280 pb when cuts on the q of |y| < 5 and $P_T$ > 30 GeV are imposed.

For the topology of Z plus two jets two possibilities were looked at. First g + g -> q + Q + Z, or FSR of a Z.  Requiring all the final state particles to have $P_T$ > 30 GeV and |y| < 5, the cross section is about 170 pb.  Therefore the two jet plus MET > 30 GeV cross section due to this process is approximately 34 pb. Again, the dimuon events from that Z decay mode can be used to monitor and validate this contribution, albeit always with reduced statistical power. A process with a larger cross section is the "Compton scattering" with FSR of an additional gluon, q + g -> Z + q + g. The cross section requiring all the final state particles to have |y| < 5 and $P_T$ > 30 GeV is about 1.3 nb, leading to a two jet plus MET > 30 GeV cross section of about 260 pb.

The processes using valence quarks rather than sea gluons thus appear to produce larger cross sections and higher masses, Therefore, for the topology of Z plus three jets only the process q + g -> Z + q + g + g was considered. The MET cross section, again requiring all final state particles to have $|y| < 5$ and $P_T > 30$ GeV, is 32 pb. In comparison, the four jet plus MET topology has a contribution from top pairs of about 120 pb, but there is an additional charged lepton from the W decay which may be observed.

Some of the spectra leading to jets plus MET topologies are shown in Fig.14 and Fig.15. The mass spectrum will be dominated at low mass by QCD multi-jet background because of jet energy mismeasures, as discussed earlier. The Z plus jets with subsequent Z decay into neutrino pairs is irreducible but the Z decays into di-leptons can be used to establish the contribution of Z plus jets to the MET plus jets rate.

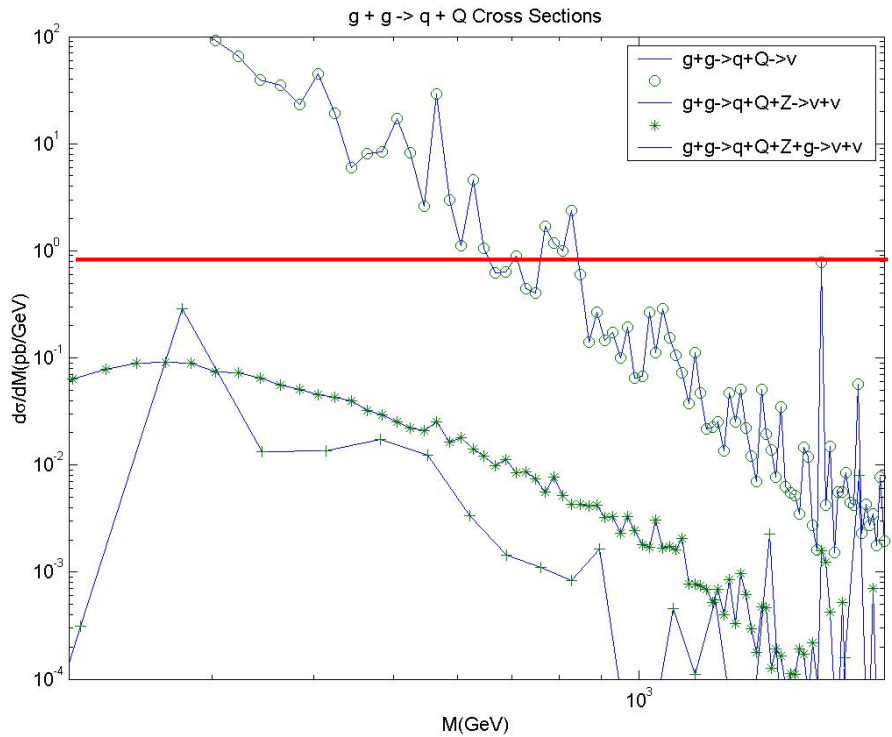

Figure 14: Mass spectrum of the final state for events with neutrinos in the final state. The cross sections shown are for b pairs with subsequent decay to neutrinos, and higher multiplicity g + g processes with the FSR of a Z or a Z and a g. The line indicates a cross section level of 1 pb/GeV.

The "Compton scattering" plus FSR processes have substantial cross sections at high mass, as seen in Fig.15. In particular the two jet plus MET and the three jet plus MET processes have rates comparable to those for the production of gluino pairs in the MMSM.

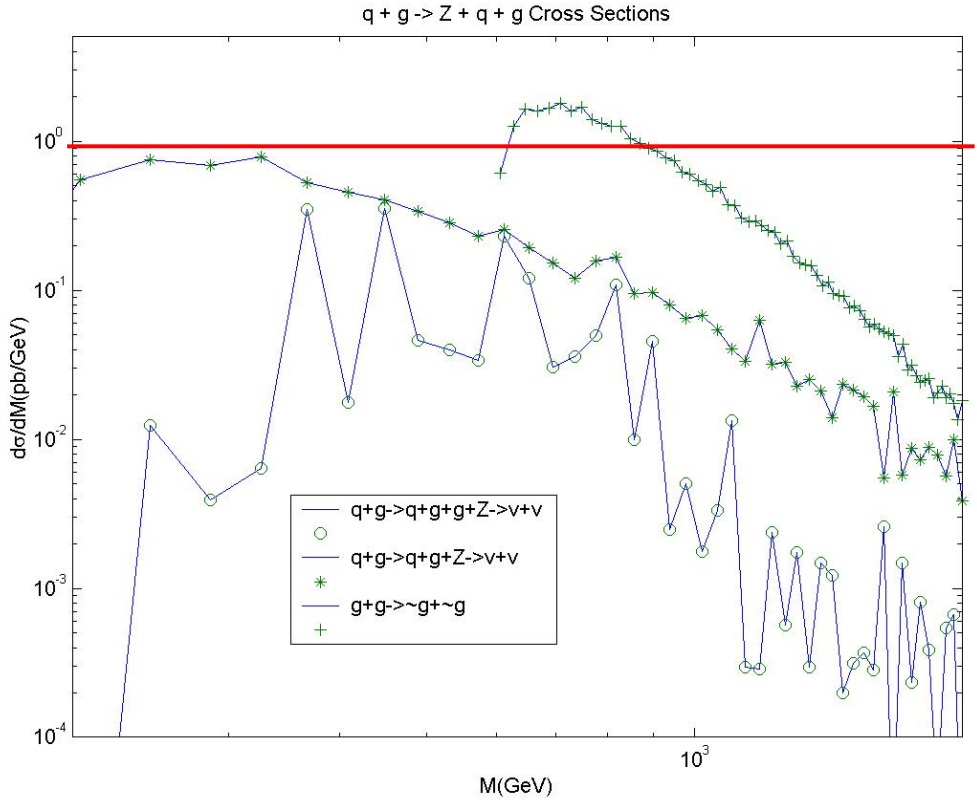

Figure 15: Cross section for Z plus jets processes initiated by g + q – i.e. using valence quarks. The basic two body process is the "Compton scattering", q +g -> Z + q. The processes displayed are the basic process with a FSR of one or two gluons. The line indicates a level of 1 pb/GeV . Also shown is the mass spectrum of a gluino pair produced by initial state gluons when the gluino mass is 300 GeV in the MMSM.

The cross section for producing top pair with a 1 TeV mass is about 0.1 pb/GeV. If one W decays into e + v, the cross section for four jets plus electron plus MET is about 0.02 pb/GeV. As seen in Fig.15, the cross section for three jets plus MET is about 0.01 pb/GeV, while that for two jets plus MET is about 0.1 pb/GeV.

Clearly, the searches for new physics in jets and MET will need to be carefully undertaken. Besides instrumental effects there are many irreducible backgrounds that need to be well understood [7]. The focus here has been on Z plus jets because that background can be quantified using the measurements of the di-lepton Z decays.

At the end of the 2008 "First Physics Run" there will have been six hundred million b pairs produced, with one hundred twenty million muons detected. These muons are sufficient in number to set the muon chamber in situ alignment. There will also be ten million W leptonic decays which can set the cross section scale. It is possible to have logged one and one half million Z decays into two muons. These events are a redundant luminosity check with the W events. They can also be used to check the lepton momentum resolution because the Z mass scale and width are well known. Having confirmed the basic SM processes; di-jets, heavy flavors, single photons, W, Z, top quark pairs, and di-photons one can with some confidence begin to search for physics beyond the SM.

There will be tens of events in excess of the present Tevatron mass reach for di-muons, di-electrons and di-photons. There will be new territory to explore in the Drell-Yan tails of the W and Z. Finally, there will have been logged potentially one hundred and twenty thousand top pair events where one W decays into muon plus neutrino while the other decays into quark pairs. In these events one can impose the precisely known W mass constraint to set the jet energy scale. Finally, if the MET plus jets signature can be well understood, the search for SUSY in these modes can be begun, searching for masses well in excess of those available at the Tevatron.

SUSY searches can also be initiated in the multi-lepton final state or in the photons plus MET final state.

At an instantaneous luminosity of $10^{33}$/cm$^2$*sec the rate of W -> $\mu$ + $\nu$ is 20 Hz, Z -> $\mu^+$ + $\mu^-$ is 3 Hz and for top pairs 0.6 Hz. Obviously, good trigger choices will need to be made so as to retain the physics processes of interest at this and higher luminosities.

**Summary and Conclusions:**

There have been other studies of early LHC running [8]. The purpose of this note is to explain how, at every available, non-zero, LHC luminosity there is work to do in preparation for discoveries. These enormously complicated detectors must be well understood, using processes which are theoretically well understood. One must understand the Standard Model backgrounds before having confidence that the understanding of the detector is sufficient to begin searches for physics beyond the SM.

Pre-operations can prepare the LHC experiments with initial calibrations and alignments. The first five orders of magnitude, up to a luminosity of $10^{27}$ /cm$^2$*sec, allow that the LHC detectors are calibrated on inelastic events and di-jets. A first look at MET is also possible, confirming that the MET distribution is due solely to jet energy mismeasurements. A di-jet mass search at masses above 2 TeV can be launched once these checks are made.

The next six orders of magnitude, up to $10^{33}$/cm$^2$*sec, or 1/10 of design luminosity allow for the launching of several additional searches. Lepton triggers can be set up. Standard candles – W and Z production – will supply a luminosity determination and a check on the lepton measurements by using the Z mass and width to set the lepton momentum scale and momentum resolution. The jet mass scale will be known using lepton plus jets triggers and the logging of many top pair events. The di-jet mass constraint using the well determined W mass will set the jet energy scale. Searches for leptons plus MET, di-leptons and di-photons can be launched into new territory beyond the reach of the Tevatron. Finally, the jets plus MET signature will be thoroughly explored. The irreducible backgrounds due to Z decays into neutrinos can be inferred using the di-lepton triggers to compared visible Z decays into lepton pairs to invisible decays into neutrino pairs.

At the end of this many order of magnitude in luminosity preparation, the LHC experiments will be well and truly launched on their voyages of discovery.